\documentstyle[12pt,epsfig,cite,amsmath,amsfonts,amssymb]{article}
\newcommand{\Prob}{\textrm{Pr}}
\newcommand{\beq}{\begin{equation}}
\newcommand{\enq}{\end{equation}}
\newcommand{\beqa}{\begin{eqnarray}}
\newcommand{\enqa}{\end{eqnarray}}
\newcommand{\beqn}{\begin{eqnarray*}}
\newcommand{\enqn}{\end{eqnarray*}}
\newcommand{\ud}{{\mathrm d}}
\newcommand{\Ei}{\textrm{Ei}}
\newcommand{\no}{\nonumber}
\newcommand{\nd}{n_1}
\newcommand{\xn}{X^{n_1}}
\newcommand{\zn}{Z^{n_1}}

\newcommand{\sxn}{x^{n_1}}
\newcommand{\szn}{z^{n_1}}
\newcommand{\gwbar}{\overline{\gamma}_E}
\newcommand{\gmbar}{\overline{\gamma}_M}

\newtheorem{theorem}{Theorem}

\newenvironment{proof}{{\sl Proof\/}:\ \ }{\qed\vspace{\baselineskip}}
\newcommand{\qed}{\hfill $\Box$}

\begin{document}

\title{On the Secrecy Capacity of Fading Channels}
\author{Praveen Kumar Gopala, Lifeng Lai and Hesham El Gamal\\
Department of Electrical and Computer Engineering\\
The Ohio State University\\
Columbus, OH 43210 \\
Email: \{gopalap,lail,helgamal\}@ece.osu.edu }
\date{}
\maketitle

\begin{abstract}
We consider the secure transmission of information over an ergodic
fading channel in the presence of an eavesdropper. Our
eavesdropper can be viewed as the wireless counterpart of Wyner's
wiretapper. The secrecy capacity of such a system is characterized
under the assumption of asymptotically long coherence intervals.
We first consider the full Channel State Information (CSI) case,
where the transmitter has access to the channel gains of the
legitimate receiver and the eavesdropper. The secrecy capacity
under this full CSI assumption serves as an upper bound for the
secrecy capacity when only the CSI of the legitimate receiver is
known at the transmitter, which is characterized next. In each
scenario, the perfect secrecy capacity is obtained along with the
optimal power and rate allocation strategies. We then propose a
low-complexity on/off power allocation strategy that achieves
near-optimal performance with only the main channel CSI. More
specifically, this scheme is shown to be asymptotically optimal as
the average SNR goes to infinity, and interestingly, is shown to
attain the secrecy capacity under the full CSI assumption.
Remarkably, our results reveal the positive impact of fading on
the secrecy capacity and establish the critical role of rate
adaptation, based on the main channel CSI, in facilitating secure
communications over slow fading channels.
\end{abstract}

\section{Introduction}
The notion of information-theoretic secrecy was first introduced
by Shannon \cite{Shan}. This strong notion of secrecy does not
rely on any assumptions on the computational resources of the
eavesdropper. More specifically, perfect information-theoretic
secrecy requires that $I(W;Z)=0$, i.e., the signal $Z$ received by
the eavesdropper does not provide any additional information about
the transmitted message $W$. Shannon considered a scenario where
both the legitimate receiver and the eavesdropper have direct
access to the transmitted signal. Under this model, he proved a
negative result implying that the achievability of perfect secrecy
requires the entropy of the private key $K$, used to encrypt the
message $W$, to be larger than or equal to the entropy of the
message itself (i.e., $H(K) \ge H(W)$ for perfect secrecy).
However, it was later shown by Wyner in \cite{Wyner} that this
negative result was a consequence of the over-restrictive model
used in~\cite{Shan}. Wyner introduced the wiretap channel which
accounts for the difference in the two noise processes, as
observed by the destination and wiretapper. In this model, the
wiretapper has no computational limitations and is assumed to know
the codebook used by the transmitter. Under the assumption that
the wiretapper's signal is a degraded version of the destination's
signal, Wyner characterized the tradeoff between the information
rate to the destination and the level of ignorance at the wiretapper
(measured by its equivocation), and showed that it is possible to
achieve a non-zero secrecy capacity. This work was later extended
to non-degraded channels by Csisz$\acute{a}$r and K\"{o}rner
\cite{Csis}, where it was shown that if the main channel is less
noisy or more capable than the wiretapper channel, then it is
possible to achieve a non-zero secrecy capacity.

More recently, the effect of slow fading on the secrecy capacity
was studied in \cite{Blahut,Barros}. In these works, it is assumed
that the fading is quasi-static which leads to an alternative
definition of outage probability, wherein secure communications
can be guaranteed only for the fraction of time when the main
channel is stronger than the channel seen by the eavesdropper.
This performance metric appears to have an operational
significance only in delay sensitive applications with full
Channel State Information (CSI). The absence of CSI sheds doubt on
the operational significance of outage-based secrecy since it
limits the ability of the source to know which parts of the
message are decoded by the eavesdropper. In this paper, we focus
on delay-tolerant applications which allow for the adoption of an
ergodic version of the slow fading channel, instead of the
outage-based formulation.
%consider the ergodic fading model,
%where the fading coefficients change within the transmission of a
%codeword. For communication systems that are delay-tolerant, the
%codeword can be made long enough to justify this assumption.
Quite interestingly, we show in the sequel that, under this model, one
can achieve a perfectly secure non-zero rate even when the eavesdropper
channel is more capable then the legitimate channel {\bf on the average}.
In particular, our work here characterizes the secrecy capacity of the
slow fading channel in the presence of an eavesdropper \cite{ICASSP}.
Our eavesdropper is the wireless counterpart of Wyner's wiretapper. We
first assume that the transmitter knows the CSI of both the legitimate
and eavesdropper channels, and derive the optimal power allocation
strategy that achieves the secrecy capacity. Next we consider the
case where the transmitter only knows the legitimate channel CSI
and, again, derive the optimal power allocation strategy.
We then propose an on/off power transmission
scheme, with variable rate allocation, which approaches the
optimal performance for asymptotically large average SNR.
Interestingly, this scheme is also shown to attain the secrecy
capacity under the full CSI assumption which implies that, at high
SNR values, the additional knowledge of the eavesdropper CSI does
not yield any gains in terms of the secrecy capacity for slow fading
channels. Finally, our theoretical and numerical results are used to
argue that rate adaptation plays a more critical role than power
control in achieving the secrecy capacity of slow fading channels. This
observation contrasts the scenario without secrecy constraints, where
transmission strategies with constant rate are able to achieve
capacity~\cite{Caire}.

\section{System Model}
\begin{figure}
\centering
\includegraphics[width=0.6\textwidth]{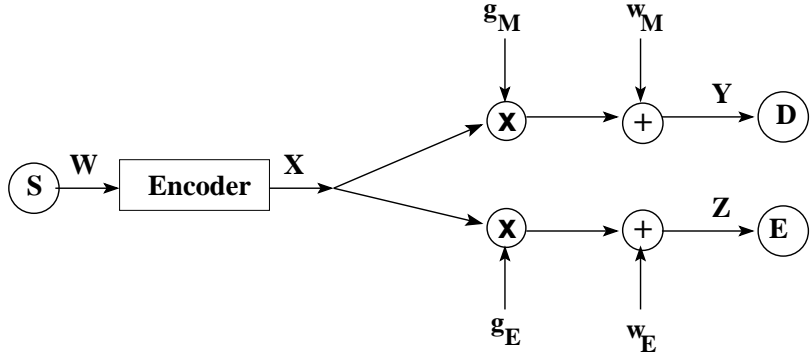} \hspace{0.5in}
\caption{The Fading Channel with an Eavesdropper\label{model}}
\end{figure}

The system model is illustrated in Fig.~\ref{model}. The source
$S$ communicates with a destination $D$ in the presence of an
eavesdropper $E$. During any coherence interval $i$, the signal received by
the destination and the eavesdropper are given by, respectively
\begin{eqnarray}
y(i)&=&g_{M}(i)x(i)+w_{M}(i),\no\\
z(i)&=&g_{E}(i)x(i)+w_{E}(i),\no
\end{eqnarray}
where $g_{M}(i),g_{E}(i)$ are the channel gains from the source to
the legitimate receiver (main channel) and the eavesdropper
(eavesdropper channel) respectively, and $w_{M}(i),w_{E}(i)$
represent the i.i.d additive Gaussian noise with unit variance at
the destination and the eavesdropper respectively. We denote the
fading power gains of the main and eavesdropper channels by
$h_M(i) = |g_M(i)|^2$ and $h_E(i) = |g_E(i)|^2$ respectively. We
assume that both channels experience block fading, where the
channel gains remain constant during each coherence interval and
change independently from one coherence interval to the next. The
fading process is assumed to be ergodic with a bounded continuous
distribution. Moreover, the fading coefficients of the destination
and the eavesdropper in any coherence interval are assumed to be
independent of each other. We further assume that the number of
channel uses $n_1$ within each coherence interval is large enough
to allow for invoking random coding arguments. As shown in the
sequel, this assumption is instrumental in our achievability
proofs.

The source wishes to send a message $W \in {\mathcal
W}=\{1,2,\cdots ,M\}$ to the destination. An $(M,n)$ code consists
of the following elements: 1) a stochastic encoder $f_n(.)$ at the
source that maps the message\footnote{The realizations of the random
variables $W,X,Y,Z$ are represented by $w,x,y,z$ respectively in
the sequel.} $w$ to a codeword $x^n \in
{\mathcal X}^{n}$, and 2) a decoding function $\phi$: ${\mathcal
Y}^{n}\rightarrow {\mathcal W}$ at the legitimate receiver. The
average error probability of an $(M,n)$ code at the legitimate
receiver is defined as \beq P_{e}^{n}~=~ \sum \limits_{w \in
{\mathcal W}} ~\frac{1}{M} \Prob (\phi (y^n)\neq w | w
\mbox{  was sent}). \enq The equivocation rate $R_e$ at the
eavesdropper is defined as the entropy rate of the transmitted
message conditioned on the available CSI and the channel outputs
at the eavesdropper, i.e., \beq R_e ~\overset{\Delta}{=}~
\frac{1}{n} H(W|Z^n,h_M^n,h_E^n) ~, \enq where $h_M^n = \{ h_M(1),
\cdots, h_M(n)\}$ and $h_E^n = \{ h_E(1), \cdots, h_E(n)\}$ denote
the channel power gains of the legitimate receiver and the
eavesdropper in $n$ coherence intervals, respectively. It
indicates the level of ignorance of the transmitted message $W$ at
the eavesdropper. In this paper we consider only perfect secrecy
which requires the equivocation rate $R_e$ to be equal to the message rate.
The perfect secrecy rate $R_s$ is said to be achievable if for any
$\epsilon>0$, there exists a sequence of codes $(2^{nR_s},n)$ such that for
any $n\geq n(\epsilon)$, we have
\beqn
&& P_{e}^{n} ~\le~ \epsilon, \\
&& R_e ~=~ \frac{1}{n} H(W|Z^n,h_M^n,h_E^n) ~\geq~ R_{s}-\epsilon. \enqn
The secrecy capacity $C_s$ is defined as the maximum achievable
perfect secrecy rate, i.e., \beq C_s ~\overset{\Delta}{=}~
\sup_{P_e^n \le \epsilon} ~ R_s ~. \enq

Throughout the sequel, we assume that the CSI is known at the
destination perfectly. Based on the available CSI, the transmitter
adapts its transmission power {\bf and rate} to maximize the
perfect secrecy rate subject to a long-term average power
constraint $\bar{P}$.

\section{Full CSI at the Transmitter}
Here we assume that at the beginning of each coherence interval,
the transmitter knows the channel states of the legitimate
receiver and the eavesdropper perfectly. When $h_M$ and $h_E$ are
both known at the transmitter, one would expect the optimal scheme
to allow for transmission only when $h_M
> h_E$, and to adapt the transmitted power according to the
instantaneous values of $h_M$ and $h_E$. The following result
formalizes this intuitive argument.

\begin{theorem} \label{thm1}
When the channel gains of both the legitimate receiver and the
eavesdropper are known at the transmitter, the secrecy capacity is
given by

\beq\label{scub} C_s^{(F)} ~=~ \max\limits_{P(h_M,h_E)}\int_{0}^{\infty}
\int_{h_E}^{\infty} \Big[\log \left( 1 + h_M P(h_M,h_E) \right)
-\log\left( 1 + h_E P(h_M,h_E) \right)\Big] f(h_M)f(h_E) \ud h_M
\ud h_E, \enq \beq \label{pcub} \mbox{such that   } \qquad \qquad
{\mathbb E}\{P(h_M,h_E)\} ~\leq~ \bar{P}. \enq
\end{theorem}
\begin{proof}
A detailed proof of achievability and the converse part is
provided in the Appendix. Here, we outline the scheme used in the
achievability part. In this scheme, transmission occurs only when
$h_M > h_E$, and uses the power allocation policy $P(h_M,h_E)$ that
satisfies the average power constraint (\ref{pcub}). Moreover, the
codeword rate at each instant is set to be $\log \left( 1 +
h_M P(h_M,h_E) \right)$, which varies according to the
instantaneous channel gains. The achievable perfect secrecy rate
at any instant is then given by $ [\log \left( 1 + h_M P(h_M,h_E)
\right) -\log\left( 1 + h_E P(h_M,h_E) \right)]^+. $ Averaging
over all fading realizations, we get the average achievable
perfect secrecy rate as
\begin{eqnarray}
R_s^{(F)} &=& \iint \left[ \log \left( 1 + h_M P(h_M,h_E) \right) -\log
\left( 1 + h_E P(h_M,h_E) \right) \right]^+ f(h_M)f(h_E) \ud h_M
\ud h_E\no\\
&=&\int_{0}^{\infty} \int_{h_E}^{\infty} \Big[\log \left( 1 + h_M
P(h_M,h_E) \right) -\log\left( 1 + h_E P(h_M,h_E) \right)\Big]
f(h_M)f(h_E) \ud h_M \ud h_E ~.\no
\end{eqnarray}
One can then optimize over all feasible power control policies
$P(h_M,h_E)$ to maximize the perfect secrecy rate.
\end{proof}

We now derive the optimal power allocation policy that achieves
the secrecy capacity under the full CSI assumption. It is easy to
check that the objective function is concave in $P(h_M,h_E)$, and
hence, by using the Lagrangian maximization approach for solving
(\ref{scub}), we get the following optimality condition
\[ \frac{\partial R_s^{(F)}}{\partial P(h_M,h_E)} ~=~
\frac{h_M}{1+h_M P(h_M,h_E)} - \frac{h_E}{1+h_E P(h_M,h_E)} -
\lambda ~=~ 0, \] whose solution is \beq \label{Ptval} P(h_M,h_E)
~=~ \frac{1}{2} \left[ \sqrt{ \left( \frac{1}{h_E} - \frac{1}{h_M}
\right)^2 + \frac{4}{\lambda} \left( \frac{1}{h_E} - \frac{1}{h_M}
\right)} - \left( \frac{1}{h_M} +
 \frac{1}{h_E} \right) \right]. \enq
If for some $(h_M,h_E)$, the value of $P(h_M,h_E)$ obtained from
(\ref{Ptval}) is negative, then it follows from the concavity of
the objective function w.r.t. $P(h_M,h_E)$ that the optimal value
of $P(h_M,h_E)$ is 0. Thus the optimal power allocation policy at
the transmitter is given by \beq \label{Ptvalnew} P(h_M,h_E) =
\frac{1}{2} \left[ \sqrt{ \left( \frac{1}{h_E} - \frac{1}{h_M}
\right)^2 + \frac{4}{\lambda} \left( \frac{1}{h_E} -
 \frac{1}{h_M} \right)} - \left( \frac{1}{h_M} + \frac{1}{h_E} \right)
 \right]^+, \enq
where $[x]^{+} = \max\{0,x\}$, and the parameter $\lambda$ is a
constant that satisfies the power constraint in (\ref{pcub}) with
equality. The secrecy capacity is then determined by substituting
this optimal power allocation policy for $P(h_M,h_E)$ in
(\ref{scub}).

\section{Only Main Channel CSI at the Transmitter}
In this section, we assume that at the beginning of each coherence
interval, the transmitter only knows the CSI of the main channel
(legitimate receiver).

\subsection{Optimal Power Allocation} \label{opt}
We first characterize the secrecy capacity under this scenario in the
following theorem.

\begin{theorem} \label{thm2}
When only the channel gain of the legitimate receiver is known at
the transmitter, the secrecy capacity is given by \beq
\label{scopt1} C_s^{(M)} ~=~ \max\limits_{P(h_M)} \iint \left[ \log
\left( 1 + h_M P(h_M) \right) - \log \left( 1 + h_E P(h_M) \right)
\right]^{+} f(h_M) f(h_E) \ud h_M \ud h_E ~, \enq \beq
\label{pcopt1} \mbox{such that   } \qquad \qquad {\mathbb
E}\{P(h_M)\} ~\leq~ \bar{P}. \enq
\end{theorem}
\begin{proof}
A detailed proof of achievability and the converse part is
provided in the Appendix. Here, we outline the scheme used to show
achievability. We use the following {\bf variable rate}
transmission scheme. During a coherence interval with main channel
fading state $h_M$, the transmitter transmits codewords at rate
$\log(1+h_M P(h_M))$ with power $P(h_M)$. This variable rate
scheme relies on the assumption of large coherence intervals and
ensures that when $h_E>h_M$, the mutual information between the
source and the eavesdropper is upper bounded by $\log(1+h_M
P(h_M))$. When $h_E \le h_M$, this mutual information will be
$\log(1+h_E P(h_M))$. Averaging over all the fading states, the
average rate of the main channel is given by
\[ \iint \log \left( 1 + h_M P(h_M) \right)
f(h_M) f(h_E) \ud h_M \ud h_E, \] while the information
accumulated at the eavesdropper is
\[ \iint\log \left( 1 + \min\{h_M,h_E\} P(h_M)
\right) f(h_M) f(h_E) \ud h_M \ud h_E. \] Hence for a given power
control policy $P(h_M)$, the achievable perfect secrecy rate is
given by \beq \label{sec_cap} R_s^{(M)} ~=~ \iint \left[ \log
\left( 1 + h_M P(h_M) \right) - \log \left( 1 + h_E P(h_M) \right)
\right]^{+} f(h_M) f(h_E) \ud h_M \ud h_E. \enq One can then
optimize over all feasible power control policies $P(h_M)$ to
maximize the perfect secrecy rate. Finally, we observe that our
secure message is {\bf hidden} across different fading states
(please refer to our proof for more details).
\end{proof}

We now derive the optimal power allocation policy that achieves
the secrecy capacity under the main channel CSI assumption.
Similar to Theorem~\ref{thm1}, the objective function under this
case is also concave, and using the Lagrangian maximization
approach for solving (\ref{scopt1}), we get the following
optimality condition.
\[ \frac{\partial R_s^{(M)}}{\partial P(h_M)} ~=~
\frac{h_M \Prob \left(h_E \le h_M \right)}{1+h_M P(h_M)} -
\int_{0}^{h_M} \left( \frac{h_E}{1+h_E P(h_M)} \right) f(h_E)\ud
h_E - \lambda ~=~ 0, \] where $\lambda$ is a constant that
satisfies the power constraint in (\ref{pcopt1}) with equality.
For any main channel fading state $h_M$, the optimal transmit
power level $P(h_M)$ is determined from the above equation. If the
obtained power level turns out to be negative, then the optimal
value of $P(h_M)$ is equal to 0. This follows from the concavity
of the objective function in (\ref{scopt1}) w.r.t. $P(h_M)$. The
solution to this optimization problem depends on the distributions
$f(h_{M})$ and $f(h_{E})$. In the following, we focus on the
Rayleigh fading scenario with ${\mathbb E}\{h_M\}= \gmbar$ and
${\mathbb E}\{ h_E\}=\gwbar$ in detail. With Rayleigh fading, the
objective function in (\ref{scopt1}) simplifies to \beqa C_s^{(M)}
&=&\max\limits_{P(h_M)} ~ \int_{0}^{\infty} \Big[ \left( 1 -
e^{-(h_M/\gwbar)} \right)
\log \left( 1 + h_M P(h_M) \right) - \no \\
&&  \qquad \qquad \qquad  \left. \int_{0}^{h_M} \log \left( 1 + h_E
P(h_M) \right) \frac{1}{\gwbar} e^{-(h_E/\gwbar)} \ud h_E \right]
\frac{1}{\gmbar} e^{-(h_M/\gmbar)} \ud h_M  \no \\[0.03in]
&=& \max\limits_{P(h_M)} ~ \int_{0}^{\infty} \left[ \log \left( 1 + h_M P(h_M) \right) -
\exp \left(\frac{1}{\gwbar P(h_M)}\right) \left( \Ei \left(
\frac{1}{\gwbar P(h_M)} \right) - \right. \right. \no \\
&& \hspace{2.3in} \left. \left. \Ei \left( \frac{h_M}{\gwbar} +
\frac{1}{ \gwbar P(h_M)} \right) \right) \right] \frac{1}{\gmbar}
e^{-(h_M/\gmbar)} \ud h_M , \label{scopt} \enqa where \[ \Ei (x)
~=~ \int_{x}^{\infty} \frac{e^{-t}}{t} ~\ud t ~. \hspace{2in} \]
Specializing the optimality conditions to the Rayleigh fading
scenario, it can be shown that the power level of the transmitter
at any fading state $h_M$ is obtained by solving the equation
\beqn \left( 1 - e^{-(h_M/\gwbar)} \right) \left( \frac{h_M}{1+
h_M P(h_M)} \right) - \lambda - \frac{\left( 1 - e^{-(h_M/\gwbar)}
\right)}{P(h_M)} + \qquad \qquad \qquad \qquad \qquad \\\qquad
\qquad \qquad \qquad \frac{\exp \left( \frac{1}{\gwbar P(h_M)}
\right)} {\gwbar (P(h_M))^2} \left[ \Ei \left(\frac{1}{\gwbar
P(h_M)}\right) - \Ei \left(\frac{h_M}{\gwbar} + \frac{1}{\gwbar
P(h_M)} \right) \right] ~=~ 0 . \enqn If there is no positive
solution to this equation for a particular $h_M$, then we set
$P(h_M)=0$. The secrecy capacity is then determined by
substituting this optimal power allocation policy for $P(h_M)$ in
(\ref{scopt}).

We observe that, unlike the traditional ergodic fading scenario,
achieving the optimal performance under a security constraint
relies heavily on using a variable rate transmission strategy.
This can be seen by evaluating the performance of a constant rate
strategy where a single codeword is interleaved across infinitely
many fading realizations. This interleaving will result in the
eavesdropper {\bf gaining more information}, than the destination,
when its channel is better than the main channel, thereby yielding
a perfect secrecy rate that is strictly smaller than that in
(\ref{sec_cap}). It is easy to see that the achievable perfect
secrecy rate of the constant rate scheme, assuming a Gaussian
codebook, is given by
\[ \max\limits_{P(h_M)} \iint \left[ \log \left( 1 + h_M P(h_M)
\right) - \log \left( 1 + h_E P(h_M) \right) \right] f(h_M) f(h_E)
\ud h_M \ud h_E ~, \]
\[ \mbox{such that  } \qquad \qquad {\mathbb E}\{P(h_M)\} \le \bar{P}. \]

Unlike the two previous optimization problems, the objective
function in this optimization problem is not a concave function of
$P(h_M)$. Using the Lagrangian formulation, we only get the
following {\em necessary} KKT conditions for the optimal point.

\[ P(h_M)\left[\lambda-\frac{h_M}{1+h_MP(h_M)}+\int \left( \frac{h_E}{1+h_EP(h_M)}
\right) f(h_E) \ud h_E\right]~=~0, \]
\[ \lambda ~\geq~ \frac{h_M}{1+h_MP(h_M)}-\int \left( \frac{h_E}{1+h_EP(h_M)} \right) f(h_E) \ud h_E, \]
\beq {\mathbb E} \{P(h_M)\} ~=~ \bar{P}. \enq

\subsection{On/Off Power Control}
We now propose a transmission policy wherein the transmitter sends
information only when the channel gain of the legitimate receiver
$h_M$ exceeds a pre-determined constant threshold $\tau > 0$.
Moreover, when $h_M>\tau$, the transmitter always uses the same
power level $P$. However, it is crucial to adapt the rate of
transmission instantaneously as $\log(1+P h_M)$ with $h_M$. It is
clear that for an average power constraint $\bar{P}$, the constant
power level used for transmission will be
\[ P ~=~ \frac{\bar{P}}{\Prob (h_M > \tau)} ~. \]
Using a similar argument as in the achievable part of
Theorem~\ref{thm2}, we get the perfect secrecy rate achieved by the
proposed scheme, using Gaussian inputs, as
\[ R_s^{(CP)} ~=~ \int_{0}^{\infty} \int_{\tau}^{\infty} \left[ \log \left( 1 +
h_M P \right) -  \log \left( 1 + h_E P \right) \right]^{+} f(h_M)
f(h_E) \ud h_M \ud h_E ~. \]
Specializing to the Rayleigh fading scenario, we get
\[ P ~=~ \frac{\bar{P}}{\Prob (h_M > \tau)} ~=~ \bar{P} e^{(\tau/\gmbar)} ~, \] and
the secrecy capacity simplifies to
\[ R_s^{(CP)} = \int_{\tau}^{\infty} \int_{0}^{h_M}  \left[ \log \left( 1 +
h_M \bar{P} e^{(\tau/\gmbar)} \right) -  \log \left( 1 + h_E
\bar{P} e^{(\tau/\gmbar)} \right) \right] \frac{1}{\gmbar}
e^{-(h_M/\gmbar)} \frac{1}{\gwbar} e^{-(h_E/\gwbar)} \ud h_E \ud
h_M ~,  \] \vspace{0.01in} which then simplifies to
\[  R_s^{(CP)} ~=~ e^{-(\tau/\gmbar)} \log \left( 1 + \tau \bar{P} e^{(\tau/\gmbar)}
\right) + \exp \left(\frac{1}{\gmbar \bar{P}
e^{(\tau/\gmbar)}}\right) \Ei \left( \frac{\tau}{\gmbar}
+\frac{1}{\gmbar \bar{P} e^{(\tau/\gmbar)}} \right)  \]
\[ + \exp \left( \frac{1}{\gwbar \bar{P} e^{(\tau/\gmbar)}} - \frac{\tau}{\gmbar} \right)
\left[ \Ei \left(\frac{\tau}{\gwbar} + \frac{1}{\gwbar \bar{P}
e^{(\tau/\gmbar)}} \right) - \Ei \left(\frac{1}{\gwbar \bar{P}
e^{(\tau/\gmbar)}} \right) \right] \] \[ - \exp \left(
\frac{\left[ \frac{1}{\gmbar} + \frac{1}{\gwbar} \right]}{\bar{P}
e^{(\tau/\gmbar)}}\right) \Ei \left( \left[\frac{1}{\gmbar} +
\frac{1}{\gwbar} \right] \left[\tau + \frac{1}{\bar{P}
e^{(\tau/\gmbar)}} \right] \right) ~. \] One can then optimize
over the threshold $\tau$ to get the maximum achievable perfect
secrecy rate.
%The optimal threshold $\tau^{*}$ can be calculated from the
%following equation.
%\[ e^{-\tau} \left\{ \log \left( 1 + \tau P e^{\tau} \right) + \exp \left(
%\frac{1}{P e^{\tau}}\right) \left[ \Ei \left(\tau + \frac{1}{P e^{\tau}}
%\right) - \Ei \left(\frac{1}{P e^{\tau}} \right) \right] \left[ 1 + \frac{1}{P
%e^{\tau}} \right] \right\} \]
%\[ \qquad + \left( \frac{1}{P e^{\tau}} \right) \exp \left( \frac{1}{P e^{\tau}}
%\right)  \Ei \left(\tau + \frac{1}{P e^{\tau}} \right) - \left( \frac{2}{P
%e^{\tau}} \right) \exp \left( \frac{2}{P e^{\tau}} \right)  \Ei \left(2\tau +
%\frac{2}{P e^{\tau}} \right) ~=~ 0. \]

%\subsection{Asymptotic Analysis}
Finally, we establish the asymptotic optimality of this on/off
scheme as the available average transmission power
$\bar{P}\to\infty$. For the on/off power allocation policy, we
have
\[ R_s^{(CP)} = \lim_{\bar{P} \to \infty} ~\int_{\tau^{*}}^{\infty} \int_0^{h_M} \log \left(
\frac{1+h_M P}{1+h_E P} \right) f(h_M) f(h_E) \ud h_E \ud h_M .\]
Taking $\tau^{*} =0$, we get $P=\bar{P}$ and \beqa R_s^{(CP)} &\ge&
\lim_{\bar{P} \to \infty} ~\int_{0}^{\infty} \int_0^{h_M} \log
\left(
\frac{(1/\bar{P})+h_M}{(1/\bar{P})+h_E} \right) f(h_M) f(h_E) \ud h_E \ud h_M \no \\
&\overset{(a)}{=}& \int_{0}^{\infty} \int_0^{h_M} \lim_{\bar{P}
\to \infty} ~ \log \left(
\frac{(1/\bar{P})+h_M}{(1/\bar{P})+h_E} \right) f(h_M) f(h_E) \ud h_E \ud h_M \no \\
&=&\int_{0}^{\infty} \int_0^{h_M} \log \left( \frac{h_M}{h_E}
\right) f(h_M) f(h_E) \ud h_E \ud h_M ~=~ {\mathbb E}_{\{h_M >
h_E\}}\left\{\log \left( \frac{h_M}{h_E} \right) \right\},
\label{lb} \enqa where (a) follows from the Dominated Convergence
Theorem, since
\[ \left| \log \left( \frac{(1/\bar{P})+h_M}{(1/\bar{P})+h_E} \right) \right| ~\le~
\left| \log \left( \frac{h_M}{h_E} \right) \right|, \qquad \forall
\bar{P} \quad \mbox{when $h_M > h_E$,} \]
\[ \mbox{and} \qquad \int_{0}^{\infty} \int_0^{h_M} \log \left( \frac{h_M}{h_E}
\right) f(h_M) f(h_E) \ud h_E \ud h_M  < \infty, \] since
${\mathbb E}\{h_M\} <\infty$, $\Big| \int_0^1 \log x ~\ud x \Big| = 1 < \infty$ and
$f(h_M),f(h_E)$ are continuous and bounded.

Now under the full CSI assumption, we have \beq C_s^{(F)} ~=~ {\mathbb
E}_{\{h_M > h_E\}}\left\{\log \left( \frac{\frac{1}{P(h_M,
h_E)}+h_M}{\frac{1}{P(h_M,h_E)}+h_E} \right) \right\} ~\le~
{\mathbb E}_{\{h_M > h_E\}} \left\{\log \left( \frac{h_M}{h_E}
\right) \right\} . \label{ub} \enq From (\ref{lb}) and (\ref{ub}),
it is clear that the proposed on/off power allocation policy that
uses only the main channel CSI achieves the secrecy capacity under
the full CSI assumption as $\bar{P} \to \infty$. Thus the absence
of eavesdropper CSI at the transmitter does not reduce the secrecy
capacity at high SNR values.

%\section{Only Receiver CSI}

\section{Numerical Results}
\begin{figure}
\centering
\includegraphics[width=0.8\textwidth,height=0.45\textheight]{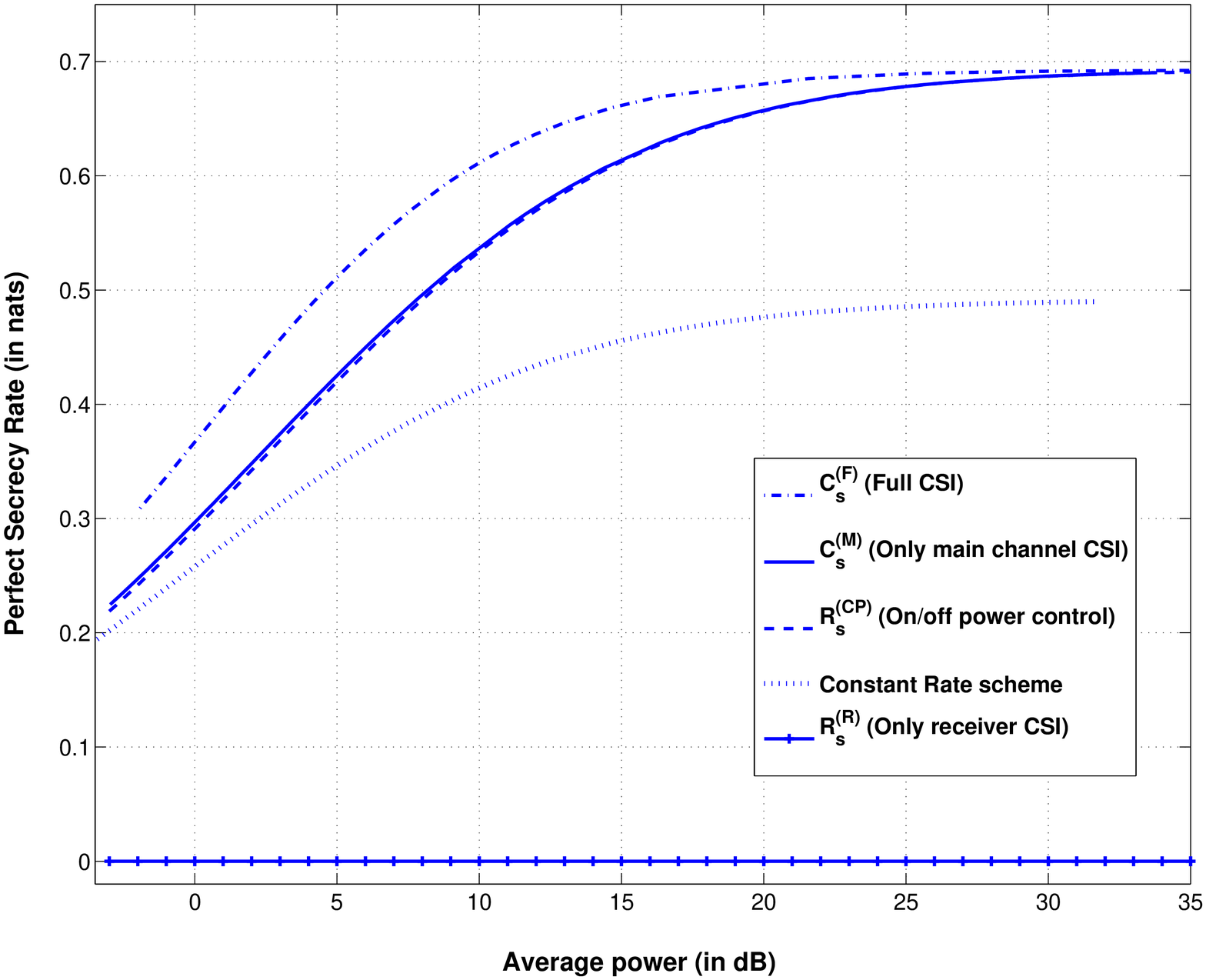}
\caption{Performance comparison for the symmetric scenario
$\gmbar = \gwbar =1$. \label{comp}}
\end{figure}

\begin{figure}
\centering
\includegraphics[width=0.8\textwidth,height=0.45\textheight]{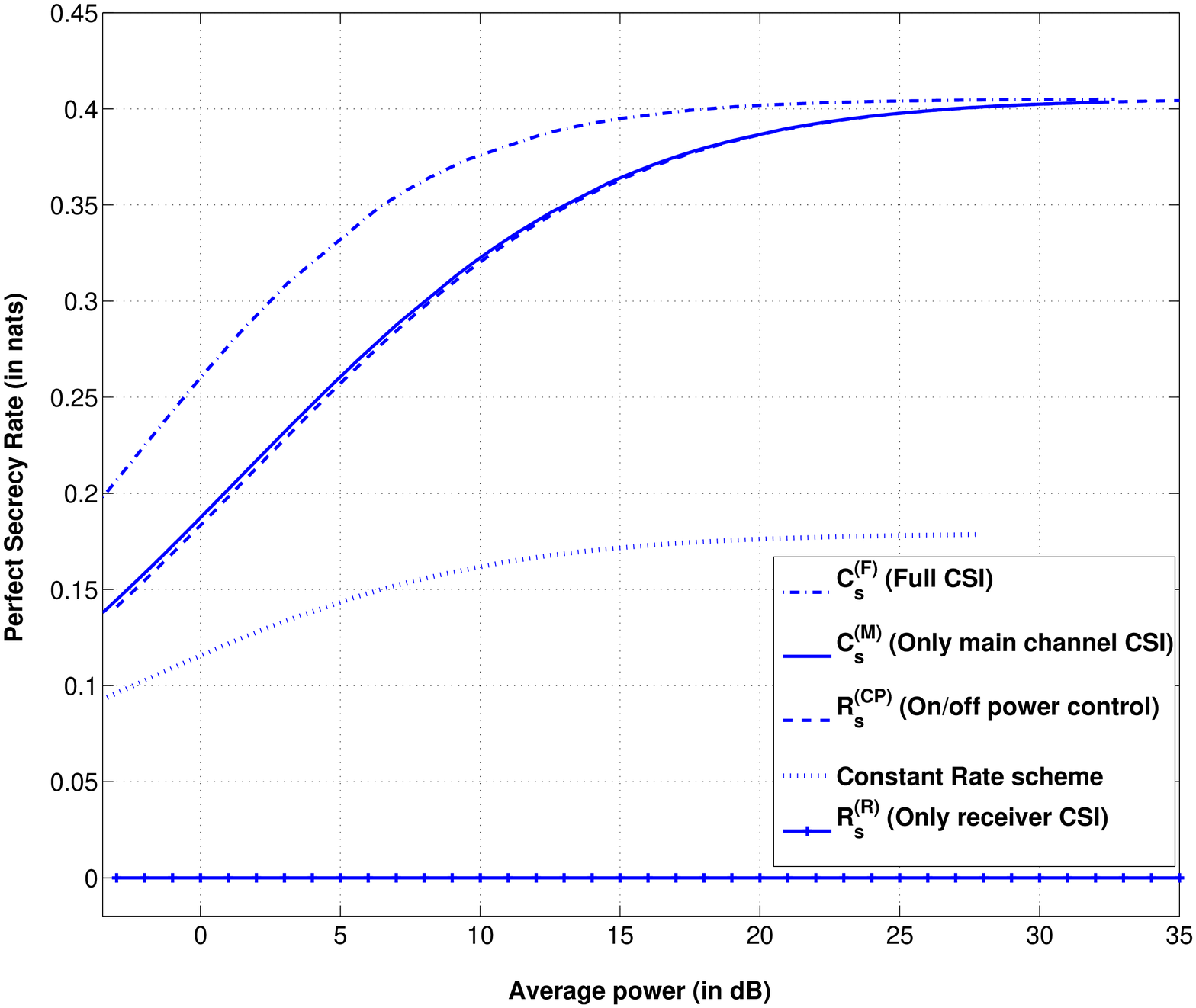}
\caption{Performance comparison for the asymmetric scenario
$\gmbar = 1$ and $\gwbar = 2$. \label{comp1}}
\end{figure}

As an additional benchmark, we first obtain the performance when
the transmitter does not have any knowledge of both the main and
eavesdropper channels (only receiver CSI). In this scenario, the
transmitter is unable to exploit rate/power adaptation and always
transmits with power $\bar{P}$. It is straightforward to see that
the achievable perfect secrecy rate in this scenario (using Gaussian
inputs) is given by
\beqn R_s^{(R)} &=&
\left[ \int_0^{\infty} \int_0^{\infty} \left[ \log \left( 1 + h_M
\bar{P} \right)
-  \log \left( 1 + h_E \bar{P} \right) \right] f(h_M) f(h_E) \ud h_M \ud h_E \right]^{+}  \\
&=& \left[ \int_0^{\infty} \log \left( 1 + h_M \bar{P} \right)
f(h_M) \ud h_M  -
 \int_0^{\infty} \log \left( 1 + h_E \bar{P} \right) f(h_E) \ud h_E \right]^{+} ~, \enqn
which reduces to the following for the Rayleigh fading scenario
\[ R_s^{(R)} ~=~ \left[ \exp \left( \frac{1}{\gmbar \bar{P}} \right) \Ei \left(
\frac{1}{\gmbar \bar{P}} \right) - \exp \left( \frac{1}{\gwbar
\bar{P}} \right) \Ei \left( \frac{1}{\gwbar \bar{P}} \right)
\right]^{+} . \] Thus when $\gwbar \ge \gmbar$, $R_s^{(R)} = 0$. The
results for the Rayleigh normalized-symmetric case ($\gmbar =
\gwbar =1$) are presented in Fig.~\ref{comp}. It is clear that the
performance of the on/off power control scheme is very close to
the secrecy capacity (with only main channel CSI) for a wide range
of SNRs and, as expected, approaches the secrecy capacities, under
both the full CSI and main channel CSI assumptions, at high values
of SNR. The performance of the constant rate scheme is much worse
than the other schemes that employ rate adaptation. Here we note
that the performance curve for the constant rate scheme might be a
lower bound to the secrecy capacity (since the KKT conditions
are necessary but not sufficient for non-convex optimization).
We then consider an asymmetric scenario, wherein the
eavesdropper channel is more capable than the main channel, with
$\gmbar=1$ and $\gwbar=2$. The performance results for this
scenario are plotted in Fig.~\ref{comp1}. Again it is clear from
the plot that the performance of the on/off power control scheme
is optimal at high values of SNR, and that rate adaptation schemes
yield higher perfect secrecy rates than constant rate transmission
schemes.
%{\bf Need to add results for the
%constant rate scheme. Please add also a curve for the performance
%with only receiver CSI}

\section{Conclusions}
We have characterized the secrecy capacity of the slow fading channel
with an eavesdropper under different assumptions on the available
transmitter CSI. Our work establishes the interesting result that
a non-zero perfectly secure rate is achievable in the fading
channel even when the eavesdropper is more capable than the
legitimate receiver (on the average). By contrasting this
conclusion with the traditional AWGN scenario, one can see the
positive impact of fading on {\bf enhancing} the secrecy capacity.
Furthermore, we proposed a low-complexity on/off power
transmission scheme and established its asymptotic optimality.
This optimality shows that the presence of eavesdropper CSI at the
transmitter does not offer additional gains in the secrecy
capacity for slow fading channels, at high enough SNR levels.
The knowledge of the main channel CSI, however, is crucial since
it is easy to see that the absence of this information leads to a
zero secrecy capacity when the eavesdropper is more capable than
the legitimate receiver on the average. Finally, our theoretical
and numerical results established the critical role of appropriate
rate adaptation in facilitating secure communications over slow fading
channels.

\appendix

\section{Proof of Theorem~\ref{thm1}}
We first prove the achievability of (\ref{scub}) by showing that
for any perfect secrecy rate $R_s < C_s^{(F)}$, there exists a sequence of
$(2^{n R_s},n)$ block codes with average power $\bar{P}$,
equivocation rate $R_e > R_s - \epsilon$, and probability of error $P_e^n \to
0$ as $n \to \infty$. Let $R_s = C_s^{(F)} - 3 \delta$ for some $\delta
>0$. We quantize the main channel gains $h_M \in [0,M_1]$ into
uniform bins $\{h_{M,i}\}_{ i=1}^{q_1}$, and the eavesdropper
channel gains $h_E \in [0,M_2]$ into uniform bins
$\{h_{E,j}\}_{j=1}^{q_2}$. The channels are said to be in state
$s_{ij}$ ($i \in [1,q_1]$, $j \in [1,q_2]$), if $h_{M,i} \le h_M <
h_{M,(i+1)}$ and $h_{E,j} \le h_E < h_{E,(j+1)}$, where
$h_{M,(q_1+1)}=M_1, h_{E,(q_2+1)}= M_2$. We also define a power
control policy for any state $s_{ij}$ by
\beq \label{Ptdef}
P(h_{M,i},h_{E,j}) ~=~ \inf_{h_{M,i} \le h_M < h_{M,(i+1)}, h_{E,j} \le
h_E < h_{E,(j+1)}} P(h_M,h_E) ~, \enq
where $P(h_M,h_E)$ is the optimal power allocation policy in (\ref{Ptvalnew})
that satisfies $P(h_M,h_E) =0$ for all $h_M \le h_E$, and the power constraint
\beq \label{Pcon1} \int_{0}^{\infty} \int_{h_E}^{\infty}
P(h_M,h_E) f(h_M) f(h_E) \ud h_M \ud h_E ~\le~ \bar{P}. \enq
Consider a time-invariant AWGN channel with channel gains $h_M \in
[h_{M,i}, h_{M,(i+1)})$ and $h_E \in [h_{E,j}, h_{E,(j+1)})$. It is
shown in \cite{Hell,Liang} that for this channel, we can develop a
sequence of $(2^{n_{ij} (R_s)_{ij}},n_{ij})$ codes with codeword
rate $\log \left( 1 + h_{M,i} P(h_{M,i},h_{E,j}) \right)$ and perfect
secrecy rate
\beq \label{Reij} (R_s)_{ij} ~=~\Big[ \log \left(1 + h_{M,i}
P(h_{M,i},h_{E,j}) \right) - \log \left(1 + h_{E,(j+1)} P(h_{M,i},h_{E,j})
\right)\Big]^+ ~, \enq such that the average power is
$P(h_{M,i},h_{E,j})$ and with error probability $P_e^{ij} \to 0$
as $n_{ij} \to \infty$, where
\[ n_{ij} = n ~\Prob \left(h_{M,i} \le h_M < h_{M,(i+1)}, h_{E,j} \le
h_E < h_{E,(j+1)} \right) \]
for sufficiently large $n$. Note that the expression in (\ref{Reij})
is obtained by considering the worst case scenario $h_M=h_{M,i},
h_E=h_{E,(j+1)}$ that yields the smallest perfect secrecy rate.

For transmitting the message index $w \in \{1 , \cdots,2^{n R_s}
\}$, we first map $w$ to the indices $\{w_{ij}\}$ by dividing
the $n R_s$ bits which determine the message index into sets of
$n_{ij} (R_s)_{ij}$ bits. The transmitter uses a multiplexing strategy
and transmits codewords $\{x_{w_{ij}}\}$ at codeword rate $\log \left(
1 + h_{M,i} P(h_{M,i},h_{E,j}) \right)$ and perfect secrecy rate $(R_s)_{ij}$,
when the channel is in state $s_{ij}$. As $n \to \infty$, this scheme
achieves the perfect secrecy rate (using the ergodicity of the channel),
\[ R_s ~=~ \sum_{i=1}^{q_1} \sum_{j=1}^{q_2} \left[
\log \left( \frac{1 + h_{M,i} P(h_{M,i},h_{E,j})}{1 + h_{E,(j+1)} P(h_{M,i},h_{E,j})}
\right) \right]^+ \Prob \left( h_{M,i} \le h_M < h_{M,(i+1)} ,h_{E,j} \le h_E <
h_{E,(j+1)} \right) . \]
Thus for a fixed $\delta$, we can find a sufficiently large $n$
such that \beq \label{Re1}
R_s ~\ge~ \sum_{i=1}^{q_1} \sum_{j=1}^{q_2} \left[
\log \left( \frac{1 + h_{M,i} P(h_{M,i},h_{E,j})}{1 + h_{E,(j+1)}
P(h_{M,i},h_{E,j})} \right) \right]^+ \Prob \left( h_{M,i} \le h_M < h_{M,(i+1)}
,h_{E,j} \le h_E < h_{E,(j+1)} \right) - \delta . \enq
For asymptotically large $n$, using the ergodicity of the channel,
the average power of the multiplexing scheme satisfies
\beqn && \sum_{i=1}^{q_1} \sum_{j=1}^{q_2} P(h_{M,i},h_{E,j}) \int_{h_{M,i}}^{
h_{M,(i+1)}} \int_{h_{E,j}}^{h_{E,(j+1)}} f(h_M) f(h_E) \ud h_M \ud h_E \\
& \overset{(a)}{\le} & \int_{0}^{\infty} \int_{0}^{\infty} P(h_M,h_E)
f(h_M) f(h_E) \ud h_M \ud h_E ~~\overset{(b)}{\le}~~ \bar{P}, \enqn
where (a) follows from the definition of $P(h_{M,i},h_{E,j})$ in
(\ref{Ptdef}) and (b) follows from (\ref{Pcon1}). Moreover, the
error probability of the multiplexing scheme is upper bounded by
\[ P_e^n ~\le~ \sum_{i=1}^{q_1} \sum_{j=1}^{q_2} P_e^{ij} ~\to~ 0,
\quad \mbox{as $n \to \infty$.} \]
Now since \beqn C_s^{(F)} &=& \int_{0}^{\infty} \int_{0}^{\infty} \left[
\log \left( 1 + h_M P(h_M,h_E) \right) - \log \left(1 + h_E
P(h_M,h_E) \right) \right]^+ f(h_M) f(h_E) \ud h_M \ud h_E \\ &
\le & \int_{0}^{\infty} \int_{h_E}^{\infty} \log \left(
\frac{h_M}{h_E} \right) f(h_M) f(h_E) \ud h_M \ud h_E ~<~ \infty,
\enqn (because ${\mathbb E}\{h_M\} < \infty$, $\Big| \int_0^1 \log x
~\ud x \Big| = 1 < \infty$ and $f(h_M),f(h_E)$ are continuous and
bounded), there exist $M_1$ and $M_2$ for a fixed $\delta$ such that
\beqa \int_{0}^{M_1} \int_{M_2}^{\infty} \left[ \log \left( 1 +
h_M P(h_M,h_E) \right) - \log \left(1 + h_E P(h_M,h_E) \right)
\right]^+ f(h_M) f(h_E) \ud h_M \ud h_E ~<~ \frac{\delta}{3}~, && \no \\
\int_{M_1}^{\infty} \int_{0}^{M_2} \left[ \log \left( 1 +
h_M P(h_M,h_E) \right) - \log \left(1 + h_E P(h_M,h_E) \right)
\right]^+ f(h_M) f(h_E) \ud h_M \ud h_E ~<~ \frac{\delta}{3}~, && \label{bnded} \\
\int_{M_1}^{\infty} \int_{M_2}^{\infty} \left[ \log \left( 1 + h_M
P(h_M,h_E) \right) - \log \left(1 + h_E P(h_M,h_E) \right) \right]^+
f(h_M) f(h_E) \ud h_M \ud h_E ~<~ \frac{\delta}{3} ~. &&  \no \enqa
Moreover, for fixed $M_1$ and $M_2$, the dominated convergence theorem implies that
\beqa && \lim_{(q_1,q_2) \to \infty} ~\sum_{i=1}^{q_1} \sum_{j=1}^{q_2}
\left[ \log \left( \frac{1 + h_{M,i} P(h_{M,i},h_{E,j})}{1 + h_{E,(j+1)}
P(h_{M,i},h_{E,j})} \right) \right]^+ \Prob \left( h_{M,i} \le h_M < h_{M,(i+1)}
,h_{E,j} \le h_E < h_{E,(j+1)} \right) \no \\
&& = \lim_{(q_1,q_2) \to \infty} ~ \sum_{i=1}^{q_1} \sum_{j=1}^{q_2}
\int_{h_{M,i}}^{h_{M,(i+1)}} \int_{h_{E,j}}^{h_{E,(j+1)}} \left[ \log
\left( \frac{1 + h_{M,i} P(h_{M,i},h_{E,j})}{1+ h_{E,(j+1)} P(h_{M,i},h_{E,j})}
\right) \right]^+ f(h_M) f(h_E) \ud h_M \ud h_E \no \\
&& =\int_{0}^{M_1} \int_0^{M_2} \left[ \log \left( \frac{1 + h_M
P(h_M,h_E)}{1 + h_E P(h_M,h_E)} \right) \right]^+ f(h_M) f(h_E)
\ud h_M \ud h_E. \label{Re2} \enqa
Choosing $M_1,M_2$ that satisfy (\ref{bnded}) and combining
(\ref{bnded}) and (\ref{Re2}), we see that for a given $\delta$,
there exist sufficiently large $q_1,q_2$ such that
\[ \sum_{i=1}^{q_1} \sum_{j=1}^{q_2} \left[ \log \left( \frac{1 +
h_{M,i} P(h_{M,i},h_{E,j})}{1 + h_{E,(j+1)} P(h_{M,i},h_{E,j})} \right) \right]^+
\Prob \left( h_{M,i} \le h_M < h_{M,(i+1)},h_{E,j} \le h_E < h_{E,(j+1)}
\right) \]
\beq \label{Re3}
\ge \int_{0}^{\infty} \int_{0}^{\infty} \left[ \log \left( \frac{ 1 +
h_M P(h_M,h_E)}{1 + h_E P(h_M,h_E)} \right) \right]^{+} f(h_M) f(h_E)
\ud h_M \ud h_E ~-~ 2\delta. \enq
Combining (\ref{Re1}) and (\ref{Re3}), we get the desired result.

We now prove the converse part by showing that for any perfect secrecy
rate $R_s$ with equivocation rate $R_e > R_s - \epsilon$ and error
probability $P_e^{n}\rightarrow 0$ as $n\rightarrow
\infty$, there exists a power allocation policy $P(h_M,h_E)$
satisfying the average power constraint, such that
$$R_s ~\leq~ \iint \left[ \log (1+h_M P(h_M,h_E)) - \log
(1+h_E P(h_M,h_E)) \right]^{+}f(h_M)f(h_E) \ud h_M \ud h_E. $$ Consider
any sequence of $(2^{nR_s},n)$ codes with perfect secrecy rate $R_s$
and equivocation rate $R_e$, such that $R_e > R_s - \epsilon$, with
average power less than or equal to $\bar{P}$ and error probability
$P_e^n \to 0$ as $n\rightarrow \infty$.
Let $N(h_M,h_E)$ denote the number of times the channel
is in fading state ($h_M,h_E$) over the interval $[0,n]$. Also let
$P^n(h_M,h_E)={\mathbb E} \big\{ \sum_{i=1}^{n} |x_{w}(i)|^2{\mathbf
1}_{\{h_{M}(i)=h_M,h_E(i)=h_E\}}\big\}$, where $\{x_w\}$ are the
codewords corresponding to the message $w$ and the expectation is
taken over all codewords. We note that the equivocation $H(W|Z^n,h_M^n,h_E^n)$
only depends on the marginal distribution of $Z^n$, and thus does not
depend on whether $Z(i)$ is a physically or stochastically degraded version
of $Y(i)$ or vice versa. Hence we assume in the following derivation that
for any fading state, either $Z(i)$ is a physically degraded version of
$Y(i)$ or vice versa (since the noise processes are Gaussian), depending
on the instantaneous channel state. Thus we have
\beqa
nR_e &=& H(W|Z^n,h_M^n,h_E^n) \no \\
&\overset{(a)}{\le}& H(W|Z^n,h_M^n,h_E^n) - H(W|Z^n,Y^n,h_M^n,h_E^n) + n\delta_n \no \\
&=& I(W;Y^n|Z^n,h_M^n,h_E^n) + n\delta_n \no \\
&\overset{(b)}{\le}&  I(X^n;Y^n|Z^n,h_M^n,h_E^n) + n\delta_n \no \\
&=& H(Y^n|Z^n,h_M^n,h_E^n) - H(Y^n|X^n,Z^n,h_M^n,h_E^n) + n\delta_n \no \\
&=& \sum_{i=1}^n \left[ H(Y(i)|Y^{i-1},Z^n,h_M^n,h_E^n) - H(Y(i)|Y^{i-1},X^n,Z^n,h_M^n,h_E^n)\right] + n\delta_n \no \hspace{0.5in} \enqa \beqa
\hspace{0.5in} &\overset{(c)}{\le}& \sum_{i=1}^n \left[ H(Y(i)|Z(i),h_M(i),h_E(i)) - H(Y(i)|X(i),Z(i),h_M(i),h_E(i)) \right] + n\delta_n \no \\
&=& \sum_{i=1}^n I(X(i);Y(i)|Z(i),h_M(i),h_E(i)) + n\delta_n \no \\
&=& \sum_{i=1}^n \iint I(X;Y|Z,h_M,h_E){\mathbf 1}_{\{h_{M}(i)=h_M,h_E(i)=h_E\}} \ud h_M \ud h_E + n\delta_n \label{conv} \\
&=& \iint I(X;Y|Z,h_M,h_E)N(h_M,h_E) \ud h_M \ud h_E + n\delta_n \no \\
&\overset{(d)}{\le}& \iint N(h_M, h_E) \left[ \log (1+h_M
P^n(h_M,h_E)) - \log (1+h_E P^n(h_M,h_E)) \right]^{+} \ud h_M \ud h_E +
n\delta_n. \no \enqa In the above derivation, (a) follows from the
Fano inequality, (b) follows from the data processing inequality
since $W \to X^n \to (Y^n,Z^n)$ forms a Markov chain, (c) follows
from the fact that conditioning reduces entropy and from the
memoryless property of the channel, (d) follows from the fact that
given $h_M$ and $h_E$, the fading channel reduces to an AWGN
channel with channel gains $(h_M,h_E)$ and average transmission
power $P^n(h_M,h_E)$, for which
$$I(X;Y|Z,h_M,h_E) ~\le~ \left[ \log (1+h_M P^n(h_M,h_E)) - \log (1+h_E
P^n(h_M,h_E)) \right]^{+} ~,$$ as shown in \cite{Hell,Liang}.
Since the codewords satisfy the power constraint, we have
\[ \iint P^n(h_M,h_E) \left( \frac{N(h_M,h_E)}{n} \right) \ud h_M
\ud h_E ~\leq~ \bar{P}. \]
For any $h_M,h_E$ such that $f(h_M,h_E)\neq 0$, $\{P^n(h_M,h_E)\}$
are bounded sequences in $n$. Thus there exists a subsequence
that converges to a limit $P(h_M,h_E)$ as $n \rightarrow \infty$.
Since for each $n$, the power constraint is satisfied, we have
\beq
\iint P(h_M,h_E) f(h_M) f(h_E) \ud h_M \ud h_E ~\leq~ \bar{P}.
\enq
Now, we have
\[ R_{e} ~\leq~ \iint \frac{N(h_M, h_E)}{n} \left[ \log \left( \frac{
1+h_M P^n(h_M,h_E)}{1+h_E P^n(h_M,h_E)} \right) \right]^{+} \ud h_M
\ud h_E + \delta_n. \]
Taking the limit along the convergent subsequence and using the ergodicity
of the channel, we get
\[ R_{e}~\leq~ \iint \left[ \log \left( \frac{1+h_M P(h_M,h_E)}{1+h_E
P(h_M,h_E)} \right) \right]^{+} f(h_M) f(h_E) \ud h_M \ud h_E + \delta_n. \]
The claim is thus proved.

\section{Proof of Theorem~\ref{thm2}}
Let $R_s = C_s^{(M)} - \delta$ for some small $\delta >0$. Let $n=\nd m$, where
$\nd$ represents the number of symbols transmitted in each coherence
interval, and $m$ represents the number of coherence intervals over
which the message $W$ is transmitted. Let $R = {\mathbb E}\{\log
\left( 1 + h_M P(h_M) \right) \} - \epsilon$. We first generate
all binary sequences $\{ {\mathbf V} \}$ of length $n R$ and then independently
assign each of them randomly to one of $2^{n R_s}$ groups, according to a
uniform distribution. This ensures that any of the
sequences are equally likely to be within any of the groups. Each secret
message $w \in \{1, \cdots, 2^{n R_s} \}$ is then assigned a group
${\mathbf V}(w)$. To encode a particular message $w$, the stochastic
encoder randomly selects a sequence ${\mathbf v}$ from the corresponding
group ${\mathbf V}(w)$, according to a uniform distribution.
This sequence ${\mathbf v}$ consisting of $n R$ bits is then sub-divided
into independent blocks $\{{\mathbf v}(1), \cdots,{\mathbf v}(m)\}$, where
the block ${\mathbf v}(i)$ consists of $n_1 \left[ \log \left( 1 + h_M(i)
P(h_M(i)) \right) - \epsilon \right]$ bits, and is transmitted in the
$i^{th}$ coherence interval ($i \in \{1, \cdots , m\}$). As $m \to \infty$,
using the ergodicity of the channel, we have
\[ \lim_{m \to \infty} ~\sum_{i=1}^m n_1 \left[ \log \left( 1 + h_M(i)
P(h_M(i)) \right) - \epsilon \right] ~=~ n_1 m \left[ {\mathbb E} \{ \log
\left( 1 + h_M P(h_M) \right) \} - \epsilon \right] ~=~ nR. \]
We then generate i.i.d. Gaussian codebooks $\{\xn(i) : i=1,\cdots,m \}$
consisting of $2^{n_1 \left[ \log \left( 1+h_M(i) P(h_M(i))\right) - \epsilon
\right]}$ codewords,
each of length $n_1$ symbols. In the $i^{th}$ coherence interval, the
transmitter encodes the block ${\mathbf v}(i)$ into the codeword
$\sxn(i)$, which is then transmitted over the fading channel. The legitimate
receiver receives $y^{n_1}(i)$ while the eavesdropper receives $\szn(i)$ in
the $i^{th}$ coherence interval. The equivocation rate at the eavesdropper
can then be lower bounded as follows.
\beqa
n R_e &=& H(W|\zn(1),\cdots ,\zn(m),h_M^n,h_E^n) \no \\
&=& H(W,\zn(1),\cdots ,\zn(m)|h_M^n,h_E^n) - H(\zn(1),\cdots ,\zn(m)|h_M^n,h_E^n) \no \\
&=& H(W,\zn(1),\cdots ,\zn(m),\xn(1),\cdots ,\xn(m)|h_M^n,h_E^n) - H(\zn(1),\cdots ,\zn(m)|h_M^n,h_E^n) \no \\
&& \hspace{2in} - H(\xn(1),\cdots ,\xn(m)|W,\zn(1),\cdots ,\zn(m),h_M^n,h_E^n) \no \\[0.03in]
&=& H(\xn(1),\cdots ,\xn(m)|h_M^n,h_E^n) + H(W,\zn(1),\cdots ,\zn(m)| \xn(1),\cdots,\xn(m),h_M^n,h_E^n) \no \\
&& - H(\zn(1),\cdots,\zn(m)|h_M^n,h_E^n) - H(\xn(1),\cdots ,\xn(m)|W,\zn(1),\cdots,\zn(m),h_M^n,h_E^n) \no \\[0.03in]
&\ge& H(\xn(1),\cdots ,\xn(m)|h_M^n,h_E^n) + H(\zn(1),\cdots ,\zn(m)| \xn(1),\cdots,\xn(m),h_M^n,h_E^n) \no \\
&& - H(\zn(1),\cdots,\zn(m)|h_M^n,h_E^n) - H(\xn(1),\cdots ,\xn(m)|W,\zn(1),\cdots,\zn(m),h_M^n,h_E^n) \no \\[0.03in]
&=& H(\xn(1),\cdots ,\xn(m)|h_M^n,h_E^n) - I(\zn(1),\cdots ,\zn(m) ; \xn(1), \cdots,\xn(m)|h_M^n,h_E^n) \no \\
&& \hspace{2in} - H(\xn(1),\cdots ,\xn(m)|W,\zn(1),\cdots,\zn(m),h_M^n,h_E^n) \no \\[0.03in]
&=& H(\xn(1), \cdots,\xn(m) | \zn(1),\cdots ,\zn(m),h_M^n,h_E^n) \no \\
&& \hspace{2in} - H(\xn(1),\cdots ,\xn(m)|W,\zn(1),\cdots,\zn(m),h_M^n,h_E^n)\no  \\
&\overset{(a)}{=}& \sum_{i=1}^m H(\xn(i) | \zn(i),h_M(i),h_E(i)) - H(\xn(1),\cdots ,\xn(m)|
W,\zn(1),\cdots,\zn(m),h_M^n,h_E^n) \no \\
&\overset{(b)}{\ge}& \sum_{i \in {\mathcal N}_m} H(\xn(i) | \zn(i),h_M(i),h_E(i)) -
H(\xn(1),\cdots ,\xn(m)| W,\zn(1),\cdots,\zn(m),h_M^n,h_E^n) \no \\
&=& \sum_{i \in {\mathcal N}_m} \left[ H(\xn(i)|h_M(i),h_E(i)) - I(\xn(i) ; \zn(i)|h_M(i),h_E(i)) \right] \no \\
&& \hspace{2in} - H(\xn(1),\cdots ,\xn(m)| W,\zn(1),\cdots,\zn(m),h_M^n,h_E^n) \no  \\
&\ge & \sum_{i \in {\mathcal N}_m} n_1 \left[ \log \left(1 + h_M(i) P(h_M(i))
\right) - \log \left( 1 + h_E(i) P(h_M(i)) \right) - \epsilon \right]\no  \\
&& \hspace{2in} - H(\xn(1),\cdots ,\xn(m)| W,\zn(1),\cdots,\zn(m),h_M^n,h_E^n) \no \enqa \beqa
&\ge& \sum_{i=1}^{m} n_1 \left\{ \left[ \log \left(1 + h_M(i) P(h_M(i)) \right) - \log
\left( 1 + h_E(i) P(h_M(i)) \right) \right]^{+} - \epsilon \right\} \no  \\
&& \hspace{2in} - H(\xn(1),\cdots ,\xn(m)| W,\zn(1),\cdots,\zn(m),h_M^n,h_E^n) \no \\
&\overset{(c)}{=}& n C_s^{(M)} - H(\xn(1),\cdots ,\xn(m)| W,\zn(1),\cdots,\zn(m),h_M^n,h_E^n)
- n \epsilon. \label{lb1}
\enqa
In the above derivation, (a) follows from
the memoryless property of the channel and the independence of the $\xn(i)$'s,
(b) is obtained by removing all those terms which correspond to the coherence
intervals $i \notin {\mathcal N}_m$, where ${\mathcal N}_m = \left\{ i \in \{1,
\cdots,m\} : h_M(i) > h_E(i) \right\}$, and (c) follows from the ergodicity of the
channel as $m \to \infty$.

Now we show that the term $H(\xn(1),\cdots,\xn(m)|W,\zn(1),\cdots,\zn(m),h_M^n,h_E^n)$
vanishes as $m,n_1 \to \infty$ by using a list decoding argument. In this
list decoding, at coherence interval $i$, the eavesdropper first constructs
a list ${\mathcal L}_i$ such that $\sxn(i) \in {\mathcal L}_i$ if
$(\sxn(i),\szn(i))$ are jointly typical. Let ${\mathcal L}={\mathcal L}_1
\times{\mathcal L}_2\times\cdots\times{\mathcal L}_m$. Given $w$, the
eavesdropper declares that $\hat{x}^n=(\sxn(1),\cdots,\sxn(m))$ was transmitted,
if $\hat{x}^n$ is the only codeword such that $\hat{x}^n \in B(w)\cap
{\mathcal L}$, where $B(w)$ is the set of codewords corresponding to the message $w$.
If the eavesdropper finds none or more than one such sequence, then it declares an
error. Hence, there are two type of error events: 1) ${\mathcal E}_1$: the transmitted
codeword $x^n_t$ is not in ${\mathcal L}$, 2) ${\mathcal E}_2$: $\exists
x^n \neq x^n_t$ such that $x^n \in B(w)\cap{\mathcal L}$.
Thus the error probability $\Prob (\hat{x}^n \neq x^n_t )= \Prob (
{\mathcal E}_1\cup {\mathcal E}_2 ) \leq \Prob ( {\mathcal E}_1) + \Prob ({\mathcal E}_2)$.
Based on the AEP, we know that $\Prob ({\mathcal E}_1) \leq \epsilon_1$. In order to
bound $\Prob ({\mathcal E}_2)$, we first bound the size of ${\mathcal L}_i$. We let
\begin{eqnarray}
\phi_i(\sxn(i)|\szn(i)) ~=~ \left\{\begin{array}{ll}1,& \textrm{when $(\sxn(i),\szn(i))$
are jointly typical,} \\ 0,& \textrm{otherwise.}\end{array} \right.
\end{eqnarray}
Now
\begin{eqnarray}
{\mathbb E}\{\|{\mathcal L}_i\|\}&=&{\mathbb E}\left\{\sum\limits_{\sxn(i)}\phi_i(
\sxn(i)|\szn(i))\right\}\no\\
&\leq&{\mathbb E}\left\{1+\sum\limits_{\sxn(i) \neq x^{n_1}_t(i)}
\phi_i(\sxn(i)|\szn(i))\right\}\no\\
&\leq&1+\sum\limits_{\sxn(i) \neq x^{n_1}_t(i)}{\mathbb E}\left\{\phi_i(\sxn(i)|
\szn(i))\right\} \no\\
&\leq&1+2^{{n_1}\left[\log(1+h_M(i)P(h_M(i)))-\log(1+h_E(i)P(h_M(i)))-\epsilon\right]}\no\\[0.03in]
&\leq&2^{{n_1}\left(\left[\log(1+h_M(i)P(h_M(i)))-\log(1+h_E(i)P(h_M(i))) - \epsilon \right]^+
+\frac{1}{n_1} \right)} ~.
\end{eqnarray}
Hence
\begin{eqnarray}
{\mathbb E}\{\|{\mathcal L}\|\}~=~\prod\limits_{i=1}^{m} {\mathbb
E}\{\|{\mathcal L}_i\|\}~\le~2^{\sum\limits_{i=1}^m n_1\left(\left[\log(1+h_M(i)P(h_M(i)))-
\log(1+h_E(i)P(h_M(i))) - \epsilon \right]^+ + \frac{1}{n_1}\right) } ~.
\end{eqnarray}
Thus
\begin{eqnarray}
\Prob ({\mathcal E}_2 ) &\leq& {\mathbb E}\left\{\sum\limits_{x^n \in{\mathcal L}, x^n \neq x^n_t}
\Prob (x^n \in B(w)) \right\}\no\\
&\overset{(a)}\leq& {\mathbb E}\left\{\|{\mathcal
L}\|2^{-n R_s}\right\} \\
&\leq& 2^{-n R_s}2^{\sum\limits_{i=1}^m
n_1\left(\left[\log(1+h_M(i)P(h_M(i)))-\log(1+h_E(i)P(h_M(i))) - \epsilon \right]^+ + \frac{1}{n_1} \right)
}\no\\
&\leq& 2^{-n \left(R_s -\frac{1}{m}\sum\limits_{i=1}^m \left(\left[\log(1+h_M(i)P(h_M(i)))
-\log(1+h_E(i)P(h_M(i))) - \epsilon \right]^+ + \frac{1}{n_1} \right) \right)} \no \\
&=& 2^{-n \left(R_s -\frac{1}{m}\sum\limits_{i=1}^m \left(\left[\log(1+h_M(i)P(h_M(i)))
-\log(1+h_E(i)P(h_M(i))) \right]^+ + \frac{1}{n_1} \right) + \frac{|{\mathcal N}_m| \epsilon}{m} \right)}, \no
\enqa
where (a) follows from the uniform distribution of the codewords in $B(w)$. Now as $n_1 \to \infty$
and $m \to \infty$, we get
\[ \Prob ({\mathcal E}_2 ) ~\le ~ 2^{-n \left(C_s - \delta - C_s + c \epsilon \right)} ~=~ 2^{-n(c \epsilon -\delta)}, \]
where $c = \Prob ( h_M > h_E)$. Thus, by choosing $\epsilon > (\delta / c)$, the error probability
$\Prob ({\mathcal E}_2 ) \to 0$ as $n \to \infty$. Now using Fano's inequality, we get
\[ H(\xn(1), \cdots ,\xn(m)|W,\zn(1),\cdots,\zn(m),h_M^n,h_E^n) ~\leq~ n \delta_{n} \qquad
\mbox{$\to 0$} \qquad \mbox{as  $n \to \infty$}. \]
Combining this with (\ref{lb1}), we get the desired result.

For the converse part, consider any sequence of $(2^{nR_s},n)$
codes with perfect secrecy rate $R_s$ and equivocation rate $R_e$,
such that $R_e > R_s - \epsilon$, with average power less than
or equal to $\bar{P}$ and error probability $P_e^n \to 0$ as
$n\rightarrow \infty$. We follow the same steps used in the proof
of the converse in Theorem~\ref{thm1} with the only difference
that now the transmission power $P^n(.)$ only depends on $h_M$. From
(\ref{conv}), we get \beqn
nR_{e}&\leq& \sum_{i=1}^n \iint I(X;Y|Z,h_M,h_E){\mathbf 1}_{\{h_{M}(i)
=h_M,h_E(i)=h_E\}} \ud h_M \ud h_E + n\delta_n \\
&=& \iint I(X;Y|Z,h_M,h_E)N(h_M,h_E) \ud h_M \ud h_E + n\delta_n \no \\
&{\le}& \iint N(h_M, h_E) \left[ \log (1+h_M P^n(h_M)) - \log
(1+h_E P^n(h_M)) \right]^{+} \ud h_M \ud h_E + n\delta_n .\no \enqn This
follows from the fact that given $h_M$ and $h_E$, the fading
channel reduces to an AWGN channel with channel gains $(h_M,h_E)$ and
average transmission power $P^n(h_M)$, for which Gaussian inputs are
known to be optimal \cite{Hell,Liang}.

Similar to the proof of Theorem~\ref{thm1}, we take the limit over the
convergent subsequence and use the ergodicity of the channel to obtain
\begin{eqnarray}
R_{e}\leq \iint \left[ \log (1+h_M P(h_M)) - \log (1+h_E P(h_M))
\right]^{+}f(h_M)f(h_E) \ud h_M \ud h_E + \delta_n,
\end{eqnarray}
where ${\mathbb E}\{P(h_M)\}\leq \bar{P}$. The claim is thus
proved.

\end{document}